\documentclass[sigconf,review=false, nonacm, timestamp=false]{acmart}
\renewcommand\footnotetextcopyrightpermission[1]{}
\AtBeginDocument{%
  }

\setcopyright{acmlicensed}
\copyrightyear{2025}
\acmYear{2025}
\acmDOI{XXXXXXX.XXXXXXX}
\acmConference[]{}{}{}

\usepackage{svg}
\usepackage{soul}
\usepackage{graphicx}
\usepackage{hyperref}
\begin{document}

\title{Capture and Interact: Rapid 3D Object Acquisition and Rendering with Gaussian Splatting in Unity}

\author{Islomjon Shukhratov}
\email{islomjon.shukhratov@imdea.org}
\orcid{0000-0003-2507-7234}
\affiliation{%
 \institution{IMDEA Networks Institute}
 \city{Madrid}
 \state{}
 \country{Spain}}

\author{Sergey Gorinsky}
\email{sergey.gorinsky@imdea.org}
\affiliation{%
 \institution{IMDEA Networks Institute}
 \city{Madrid}
 \state{}
 \country{Spain}}

\renewcommand{\shortauthors}{Shukhratov and Gorinsky}

\begin{abstract}
  Capturing and rendering three-dimensional (3D) objects in real time remain a significant challenge, yet  hold substantial potential for applications in augmented reality, digital twin systems, remote collaboration and prototyping. We present an end-to-end pipeline that leverages 3D Gaussian Splatting (3D GS) to enable rapid acquisition and interactive rendering of real-world objects using a mobile device, cloud processing and a local computer. Users scan an object with a smartphone video, upload it for automated 3D reconstruction, and visualize it interactively in Unity at an average of 150 frames per second (fps) on a laptop. The system integrates mobile capture, cloud-based 3D GS and Unity rendering to support real-time telepresence. Our experiments show that the pipeline processes scans in approximately 10 minutes on a graphics processing unit (GPU) achieving real-time rendering on the laptop.
\end{abstract}




\begin{teaserfigure}
  \centering
  \def\svgwidth{\linewidth}
\begingroup%
  \makeatletter%
  \providecommand\color[2][]{%
    \errmessage{(Inkscape) Color is used for the text in Inkscape, but the package 'color.sty' is not loaded}%
    \renewcommand\color[2][]{}%
  }%
  \providecommand\transparent[1]{%
    \errmessage{(Inkscape) Transparency is used (non-zero) for the text in Inkscape, but the package 'transparent.sty' is not loaded}%
    \renewcommand\transparent[1]{}%
  }%
  \providecommand\rotatebox[2]{#2}%
  \newcommand*\fsize{\dimexpr\f@size pt\relax}%
  \newcommand*\lineheight[1]{\fontsize{\fsize}{#1\fsize}\selectfont}%
  \ifx\svgwidth\undefined%
    \setlength{\unitlength}{720bp}%
    \ifx\svgscale\undefined%
      \relax%
    \else%
      \setlength{\unitlength}{\unitlength * \real{\svgscale}}%
    \fi%
  \else%
    \setlength{\unitlength}{\svgwidth}%
  \fi%
  \global\let\svgwidth\undefined%
  \global\let\svgscale\undefined%
  \makeatother%
  \begin{picture}(1,0.23000054)%
    \lineheight{1}%
    \setlength\tabcolsep{0pt}%
    \put(0,0){\includegraphics[width=\unitlength,page=1]{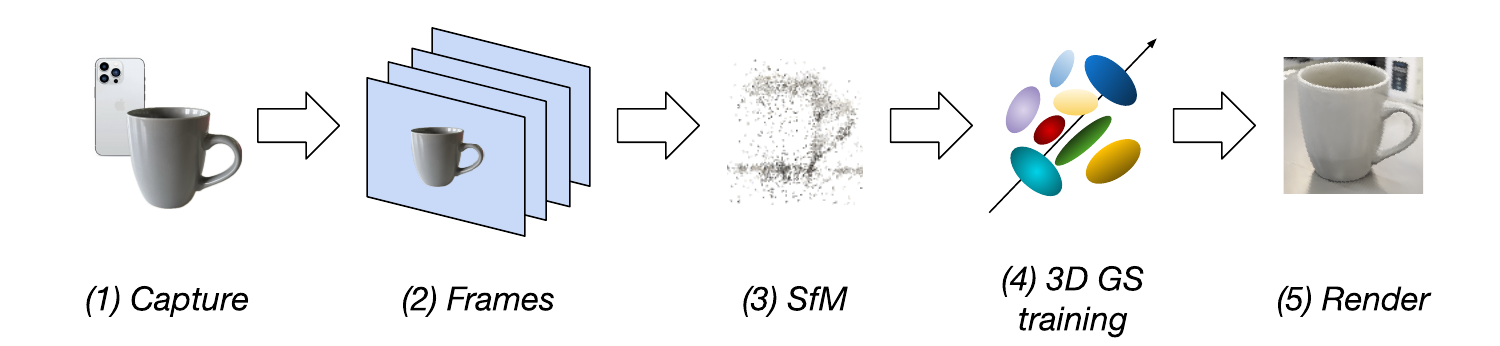}}%
  \end{picture}%
\endgroup%

  \caption{System pipeline: (1)~video capture and upload to a cloud server; (2)~video decomposition into individual frames; (3)~SfM application to generate a sparse point cloud; (4)~training of a 3D GS model on this point cloud; (5)~download of the resulting Gaussian representation to a local device and interactive rendering via Unity.}
  \label{fig:pipeline}
\end{teaserfigure}


\maketitle

\section{Introduction}

Real-time 3D object capture and rendering have a broad applications in education, virtual collaboration, healthcare and extended reality (XR) \cite{chelloug2023real}. Traditional pipelines rely on expensive hardware, controlled environments and post-processing, limiting accessibility for interactive or mobile scenarios. 

Point clouds, voxels and triangular meshes \cite{van2023tutorial} all have drawbacks: sparsity, memory intensity, high reconstruction cost and dependence on expensive hardware, such as  Light Detection and Ranging (LiDAR), with high-quality input data \cite{wang2020}. Yet, they still lack in the visual realism needed to accurately capture real-world objects.  

Neural Radiance Fields (NeRF) \cite{mildenhall2021nerf} achieve photorealistic rendering by modeling scenes as continuous functions over 3D coordinates and view directions. However, their long training times and computational intensity make them unsuitable for real-time or mobile applications such as augmented reality and live telepresence. MobileNeRF addresses this by enabling mobile rendering of pre-trained scenes using a compact rasterization-based approach, but it lacks on-device capture and end-to-end reconstruction \cite{mobilenerf}.

Karb et al. \cite{kerbl3Dgaussians} introduce 3D Gaussian Splatting (3D GS), which models the scene as a set of view-dependent Gaussians that are directly splatted onto the image plane using a fast, differentiable rasterization process. This approach enables faster training and real-time rendering performance while maintaining photorealistic quality, making it well-suited for interactive applications and real-time 3D content generation on consumer hardware.

We propose a cloud-based 3D GS pipeline that reconstructs and renders objects captured via mobile videos. The result is an interactive, photorealistic visualization, rendered in Unity on a client device. Our system demonstrates a practical approach to democratizing 3D content creation.

\section{System Overview}

The process begins with a short video recorded through a web interface on any mobile device. Once uploaded to a cloud server, the script decomposes the video into frames and processes using Structure-from-Motion (SfM) to estimate camera poses and generate a sparse point cloud.

During training, the algorithm converts each point to a 3D Gaussian, an elliptical blob defined by position, scale, rotation, opacity, color and spherical harmonic coefficients. The system applies view-dependent shading and optimizes the Gaussians over approximately 30K iterations, actively adjusting, splitting, or pruning them based on their contribution to photometric reconstruction accuracy. This results in a detailed and photorealistic 3D representation.

After optimization, the client device downloads the resulting Gaussian model and renders interactively in Unity using a real-time GPU splatting renderer. Figure~\ref{fig:pipeline} illustrates the full pipeline, from capture to interactive rendering.

\section{Implementation and Evaluation}

\begin{figure}[t!]
    \includegraphics[width=0.28\textwidth]{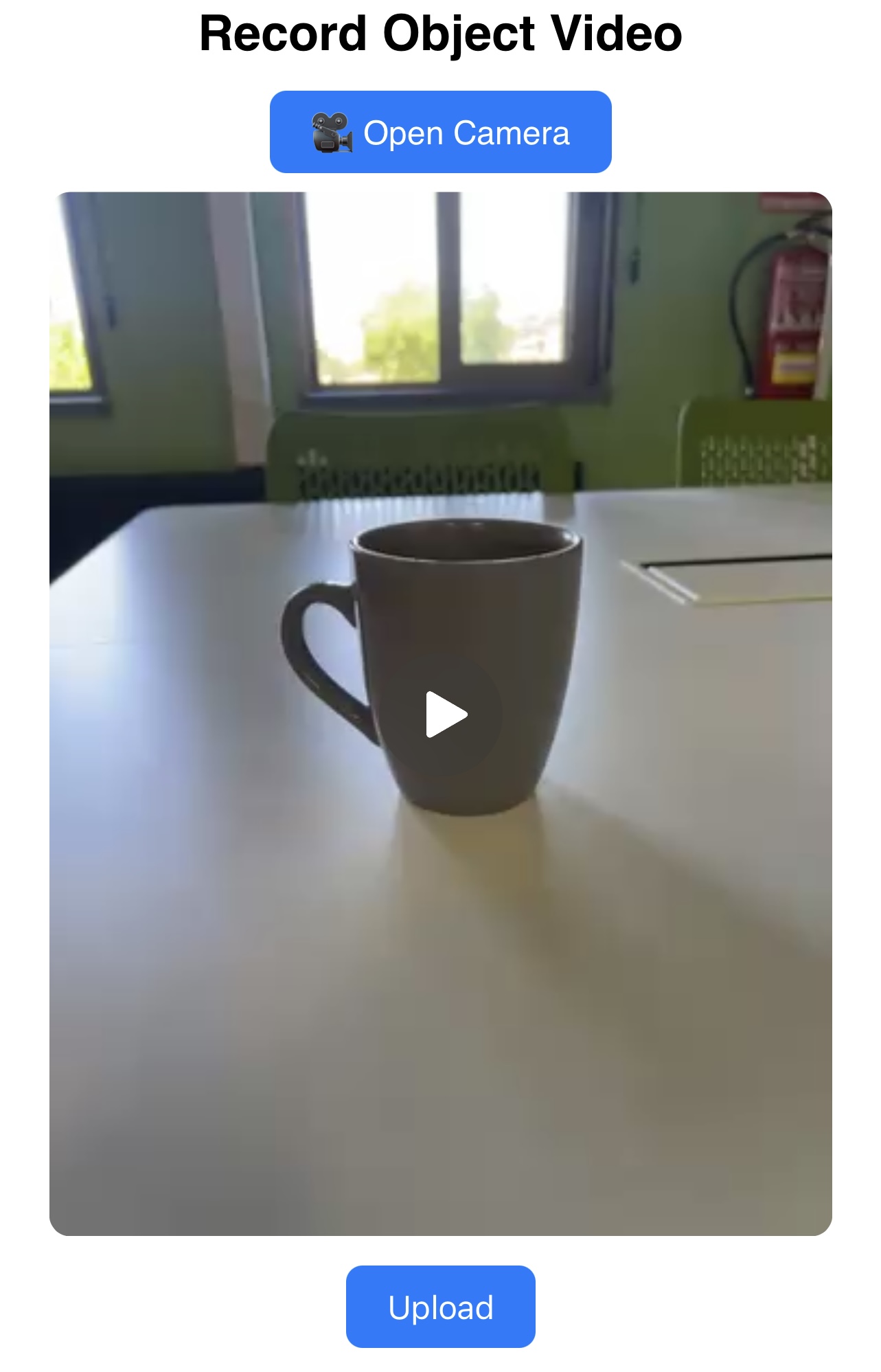}
    \caption{Web interface.}
    \label{fig:web_app}
\end{figure}

The system consists of three components: mobile video capture, cloud-based 3D GS and Unity-based real-time rendering. The mobile front-end, built with HTML5 and JavaScript, allows users to record and upload a short video (see Figure \ref{fig:web_app}). On the server side, the Structure-from-Motion in COLMAP \cite{schoenberger2016sfm, schoenberger2016mvs} extracts camera poses and reconstructs a sparse point cloud. The 3D GS pipeline relies on the open-source implementation by Kerbl et al. \cite{kerblgithub}. 

We evaluate the pipeline using both simple objects (e.g., cups and books) and complex ones (e.g., flowers, sunglasses) objects. The optimization runs for approximately 30K iterations on an NVIDIA RTX A40 GPU and the whole pipeline completes in an average of 10 minutes depending on scene complexity and video duration. On average, the reconstructed objects achieve a Peak Signal-to-Noise Ratio (PSNR) of 34.65. In future work, we plan to explore acceleration techniques for time-sensitive applications.

\begin{figure}[!b]
\centering
\begin{minipage}{0.20\textwidth}
  \centering
  
  \includegraphics[width=\linewidth]{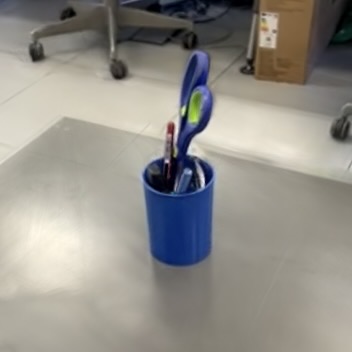}

  \includegraphics[width=\linewidth]{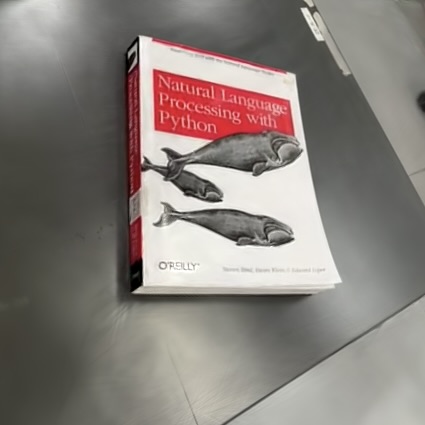}

\end{minipage}
\hfill
\begin{minipage}{0.2\textwidth}
  \centering
  \includegraphics[width=\linewidth]{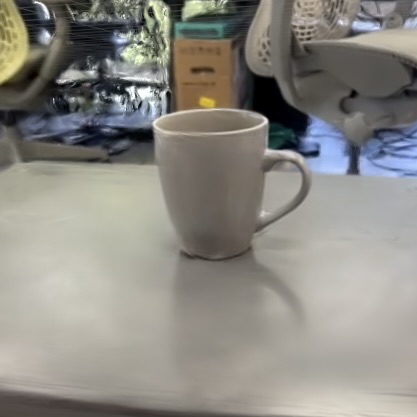}

  \includegraphics[width=\linewidth]{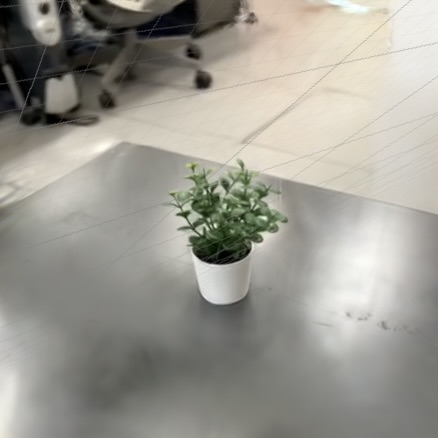}
 
\end{minipage}
\hfill

\caption{Examples of rendered objects.}
\label{fig:renders}
\end{figure}

After the client receives the resulting set of Gaussians for local rendering, a Unity-based renderer~\cite{gaussianunity} enables real-time interaction, including pan, zoom, rotation and scaling (Figure \ref{fig:renders}). The renderer operates at an average of 150~fps for approximately 500K splats on a MacBook Pro with the M4 chip.

\section{Demo Setup}

The demonstration involves a mobile device and a laptop, both connected to a cloud server. Attendees record a short video using the mobile interface. The system processes the video in the cloud and renders the reconstructed model in Unity within 10 minutes. Users then interact with the result on the local laptop. To ensure a stable and seamless demonstration, we also provide pre-captured demo objects rendered live in Unity.

\section{Conclusion and Future Work}



Our system enables rapid 3D object capture and rendering using only mobile devices, unlocking a wide range of applications across academic, educational and creative domains. Its modular and accessible design supports real-time 3D visualization. Future optimizations, such as using transformer-based feature generation instead of SfM, promise to reduce latency for time-critical use cases like live events.

The system serves diverse domains by enabling easy 3D digitization and interaction. Museums, schools and similar institutions have the potential to use it to digitize physical artifacts for documentation, virtual exhibits and hands-on learning. Researchers and teachers without 3D modeling skills create high-quality models for analysis or education. In industrial design and prototyping, users scan mockups or parts, integrate them into digital workflows and interact with them in real time, speeding up the design process.

By lowering the barrier to high-quality 3D content creation, the system encourages broader adoption of 3D GS for practical use. Future enhancements include increasing the interactivity of the viewer through integration with large language models (LLMs). For example, users gain the ability to manipulate the 3D scene intuitively using voice or text-based commands, such as "Change the color of the cup to yellow." This natural language interface enables a more accessible and hands-free experience, which proves especially valuable in educational, collaborative and assistive environments.


\bibliographystyle{ACM-Reference-Format}
\bibliography{references}

\end{document}